# Identical effects of indirect and direct electron doping of superconducting BaFe$_2$As$_2$ thin films

Takayoshi Katase[1,*], Soshi Iimura[1], Hidenori Hiramatsu[1], Toshio Kamiya[1], and Hideo Hosono[1,2,†]

[1] Materials and Structures Laboratory, Tokyo Institute of Technology, Mailbox R3-1, 4259 Nagatsuta-cho, Midori-ku, Yokohama 226-8503, Japan

[2] Frontier Research Center, Tokyo Institute of Technology, S2-6F East, Mailbox S2-13, 4259 Nagatsuta-cho, Midori-ku, Yokohama 226-8503, Japan

**Abstract**

Electron doping of a 122-type iron pnictide $BaFe_2As_2$ by substituting the Ba site with an aliovalent ion (indirect doping), which had been unsuccessful by conventional solid-state synthesis methods, was achieved by a non-equilibrium film growth process. The substitution with La was substantiated by a systematic shrinkage of the *c*-axis lattice parameter due to the smaller ionic radius of $La^{3+}$ than that of $Ba^{2+}$. A negative Hall coefficient indicated that the majority carriers were electrons, as is consistent with this aliovalent ion doping. The La substitution suppressed an antiferromagnetic transition and induced bulk superconductivity at a maximum onset critical temperature ($T_c$) of 22.4 K. The electronic phase diagram for $(Ba_{1-x}La_x)Fe_2As_2$ was built, which revealed that the indirect electron doping at the Ba site with La [$(Ba_{1-x}La_x)Fe_2As_2$] exhibits almost the same $T_c$ – doping level relation as that of the direct electron-doping at the Fe site with Co [$Ba(Fe_{1-x}Co_x)_2As_2$]. This finding clarified that $T_c$ in 122-type compounds is not affected by a crystallographic doping site, which is in sharp contrast to the 1111-type compounds, *RE*FeAsO (*RE* = rare earth). It is tentatively attributed to the differences in their dimensionality of electronic structures and electron pairing symmetries.

Since the discovery of iron pnictide superconductors[1] with high critical temperatures ($T_c$) up to 55 K,[2] there has been intensive effort to elucidate superconductivity mechanisms. Among iron pnictide superconductors, 1111-type *RE*FeAsO (*RE* = rare earth)[1] and 122-type *AE*Fe$_2$As$_2$ (*AE* = alkaline earth)[3] systems have been investigated extensively, and their electronic phase diagrams have been mapped out[4–7] to discuss superconductivity mechanisms[8–10]. Superconductivity in these compounds is induced by chemical substitution with aliovalent elements, which leads to the doping of electrons or holes into the parent materials, along with suppression of an antiferromagnetic (AFM) ordering and a tetragonal-to-orthorhombic structural phase transition[11]. The structures of these crystals are composed of alternating stacking of FeAs and *RE*O / *AE* layers, where the former forms a Fermi surface and serves as a carrier conducting path (active pairing planes in a superconducting state), while the latter acts as a nonactive blocking layer, in particular, for 1111-type compounds with thicker three-atomic *RE* oxide layers. Thus, carrier doping may be classified by crystallographic doping sites, i.e., aliovalent ion doping in the active FeAs layers ("direct doping") and in the non-active *RE*O / *AE* layers ("indirect doping") (see the inset to Fig. 4).

It is thought that direct doping causes a large deterioration on $T_c$, because impurity scattering and disorder in the active pairing layers provide a larger perturbation[12]. Indeed, it has been clarified that this is the case for the 1111-type *RE*FeAsO[5]. On the other hand, recent theoretical studies have proposed that the sensitivity to the impurity doping at Fe sites depends largely on the electron pairing symmetries[13]. Due to unique multiorbital band structures at the Fermi levels of the iron pnictides, several pairing symmetries, such as a sign-reversal $s_{+-}$ wave state[8,9] and a non-sign-reversal $s_{++}$ wave state[10], have been proposed theoretically. According to their first-principles calculation with Anderson's theorem by assuming the impurity potential for non-magnetic dopants at Fe sites[13], although the sign-reversal $s_{+-}$ wave state is sensitive to the impurity doping in the active pairing layers, the non-sign-reversal $s_{++}$ wave state is robust, i.e., $T_c$ remains almost unchanged with the doping level, opposite to the conspicuous $T_c$-drop for the $s_{+-}$ case. As noted above, it is reported that the 1111-type iron pnictides, $RE$FeAsO, exhibit pronounced differences in the maximum $T_c$ between direct electron doping [e.g.,17 K for Sm(Fe$_{1-x}$Co$_x$)AsO (Ref. 5)] and indirect electron-doping [55 K for SmFeAs(O$_{1-x}$F$_x$) (Ref. 2)]; therefore, the $s_{+-}$ wave state mediated by an AFM spin fluctuation is now accepted for the electron-pairing symmetry for the 1111-type iron pnictides. This model is supported experimentally by observation of the half-flux-quantum effect in F-doped NdFeAsO polycrystals[14]. However, for 122-type compounds, $AE$Fe$_2$As$_2$, to the best of our knowledge, this hypothesis has not been examined experimentally, because indirect electron doping has not been achieved by conventional solid-state reactions[15,16] (the only exceptions are Ref. 16 for $AE$ = Sr achieved by a high-pressure synthesis and Refs. 17 and 18 for $AE$ = Ca achieved by a melt-growth method) due to the electronic instability of the indirectly electron-doped $AE$Fe$_2$As$_2$[19].

In this Rapid Communication, we employ a nonequilibrium film-growth process to achieve metastable doping of La into the Ba sites of BaFe$_2$As$_2$. Although doping of a La ion has been unsuccessful for bulk polycrystalline samples, La dopants are successfully introduced into the BaFe$_2$As$_2$ epitaxial films grown by pulsed laser deposition (PLD). The electronic phase diagram obtained for indirect electron doping is very close to that reported for direct electron doping of Ba(Fe$_{1-x}$Co$_x$)$_2$As$_2$[7]. The close similarity between the indirect and direct doping effects in BaFe$_2$As$_2$ differs from the trends in *RE*FeAsO materials[2,5].

Here, 150–250 nm-thick films of (Ba$_{1-x}$La$_x$)Fe$_2$As$_2$ were fabricated on MgO (001) single crystals at an optimized growth temperature of 850 °C by PLD using a neodymium-doped yttrium aluminum garnet laser as the excitation source and polycrystalline bulks of La-added BaFe$_2$As$_2$ as ablation targets synthesized by a solid-state reaction. The detailed fabrication process is reported in Ref. 20. The obtained phases and their lattice parameters were examined by powder x-ray diffraction (XRD; anode radiation: CuKα, D8 ADVANCE-TXS, Bruker AXS) for polycrystalline bulk samples, while high-resolution XRD (anode radiation: CuKα$_1$, Smart Lab, RIGAKU) was used for thin films. The chemical doping concentrations of the thin films ($x_{Film}$) were estimated using a wavelength-dispersive x-ray electron-probe microanalyzer (EPMA), which also demonstrated the homogeneity of the La distribution in the (Ba$_{1-x}$La$_x$)Fe$_2$As$_2$ films (see Fig. S3 in the Supplemental Material for chemical composition mapping images). Temperature dependences of electrical resistivity ($\rho$–$T$) were measured by the four-probe method in a $T$ range of 2–300 K with a physical property measurement system (PPMS, Quantum Design). The temperature dependences of magnetization were measured with a vibrating sample magnetometer (VSM) after zero-field cooling (ZFC) and during field cooling (FC). The Hall effect measurements were performed using a six-terminal Hall bar structure formed by photolithography and Ar ion milling. The external magnetic field was applied parallel to the *c*-axis for both the magnetization and Hall effect measurements.

Figure 1 shows the evolution of the lattice parameters with doping concentration for bulk polycrystalline samples synthesized by a solid-state reaction and epitaxial films grown by PLD (see Figs. S1 and S2 in the Supplemental Material for XRD patterns). Here, the doping concentrations are taken as the nominal values for the bulk polycrystalline samples ($x_{Bulk}$) and as the chemical composition values measured by EPMA for the epitaxil films ($x_{Film}$). For the bulk polycrystals, neither the *c*- nor *a*-axis length was changed upon the impurity doping, indicating that the $La^{3+}$ ions were not incorporated into the $BaFe_2As_2$ phase by the solid-state reaction. In contrast, for the epitaxial films, the *c*-axis length systematically shrank ($\Delta c/c \sim -3.5\%$ for $x_{Film} = 0.44$), while the change in the *a*-axis was much smaller ($\Delta a/a \sim -0.3\%$). These results substantiate that $La^{3+}$ ion substitution for the $Ba^{2+}$ sites are achieved in the epitaxial films.

Figure 2 summarizes the $\rho$–$T$ curves for the $(Ba_{1-x}La_x)Fe_2As_2$ epitaxial films with $x_{Film} = 0 - 0.44$. The $\rho$ at room temperature gradually decreased with increasing $x_{Film}$. The temperature of resistivity anomaly ($T_{anom}$), which is associated with the structural and magnetic phase transitions[21], continuously shifted from 135 K to the lower *T* side as $x_{Film}$ increased. A superconducting transition with zero resistivity (an expanded view is shown in the right-hand panel of Fig. 2) was distinctly observed for $x_{Film} = 0.08$, but the resistivity anomaly still remained as observed at $T_{anom} = 72$ K, indicating that the superconductivity and the AFM ordering coexist. As $x_{Film}$ further increased, the resistivity anomaly was not detected in the $\rho$–$T$ curves. Additionally, the onset $T_c$ ($T_c^{onset}$) reached a maximum value 22.4 K at $x_{Film} = 0.13$, and then monotonically decreased to zero (normal metal state) as $x_{Film}$ further increased to $x_{Film} = 0.44$. The maximum $T_c^{onset}$ of $(Ba_{1-x}La_x)Fe_2As_2$ is comparable to $T_c^{onset}$ of 22 K for $(Sr_{1-x}La_x)Fe_2As_2$ polycrystalline samples[16], but is much lower than $T_c^{onset}$ of ~40 K for $(Ca_{1-x}RE_x)Fe_2As_2$ single crystals[17,18].

To confirm bulk superconductivity of the resulting epitaxial films, magnetization measurements were performed on the optimally-doped $(Ba_{1-x}La_x)Fe_2As_2$ films with $x_{Film}$ = 0.13 [Fig. 3(a)]. The shielding signal for ZFC was detected from $T$ = 18 K and corresponded to a 100 % shielding volume fraction at $T$ < 7 K, verifying bulk superconductivity. We also examined the effects of lattice strain because the $(Ba_{1-x}La_x)Fe_2As_2$ epitaxial films exhibited a shrinkage of lattice parameters by La doping and would be affected also by the tensile mismatch-strain from the MgO substrate. However, no significant change in $\rho$–$T$ curves was observed even after the films were subjected to post-deposition thermal annealing at 500 $^o$C for 2 h in a high-vacuum condition. Therefore, we concluded that the observed electrical properties and superconductivity in the $(Ba_{1-x}La_x)Fe_2As_2$ epitaxial films are not due to a lattice-strain effect.

In addition, to investigate the polarity of the dominant carriers, Hall effect measurements were conducted for the optimally doped $(Ba_{1-x}La_x)Fe_2As_2$ epitaxial film. Figure 3(b) shows transverse resistivity $\rho_{xy}$ in magnetic fields $\mu_0H$ up to 9 T ($\mu_0H//c$) measured at $T$ = 300–25 K. The slope $d\rho_{xy} / d(\mu_0H)$ was negative in the high $\mu_0H$ region, but a small increase in $\rho_{xy}$ was observed in the low $\mu_0H$ region. Similar to reports on $Ba(Fe_{1-x}Co_x)_2As_2$ films[22], the small increase might be due to anomalous Hall effects arising from a small magnetic impurity such as Fe and/or the $(Ba_{1-x}La_x)Fe_2As_2$ phase. The Hall coefficients ($R_H$) were estimated by excluding the anomalous $\rho_{xy}$ region at low $\mu_0H \leq 3$ T and extracting the straight line region at the higher $\mu_0H$. The inset in Fig. 3(b) shows the $T$ dependence of $R_H$. The negative $R_H$ confirmed that electrons dominate the transport properties of $(Ba_{1-x}La_x)Fe_2As_2$ films, which is consistent with the supposition that the $La^{3+}$ dopants occupy the $Ba^{2+}$ sites. $|R_H|$ gradually increased as $T$ decreased without the sudden increase associated with an AFM transition[23], and the $|R_H|$ value at 25 K ($-2.5\times10^{-3}$ cm$^3$/C) is comparable to those of optimally-doped Ba(Fe$_{1-x}$Co$_x$)$_2$As$_2$ single crystals [($-2$ to $-3$)$\times10^{-3}$ cm$^3$/C][23], which indicates that the carrier concentrations in the present (Ba$_{1-x}$La$_x$)Fe$_2$As$_2$ films are almost the same as the optimum concentration in the Ba(Fe$_{1-x}$Co$_x$)$_2$As$_2$ single crystals.

Figure 4 summarizes the electronic phase diagram of $T_{anom}$ and $T_c^{onset}$ for the (Ba$_{1-x}$La$_x$)Fe$_2$As$_2$ epitaxial films as a function of doping concentration. The previously reported phase diagram of directly electron-doped Ba(Fe$_{1-x}$Co$_x$)$_2$As$_2$ single crystals[7] is superimposed in the figure. Here the doping concentrations are normalized as the doped carriers per Fe (i.e., $x_{Film}$/2) to compare with the phase diagram of direct electron-doped Ba(Fe$_{1-x}$Co$_x$)$_2$As$_2$. For the indirect electron-doped (Ba$_{1-x}$La$_x$)Fe$_2$As$_2$ films, $T_{anom}$ rapidly decreased, and the plots of $T_c^{onset}$ formed a bell-shaped dome. The superconductivity and the AFM state coexisted only in the low doping region ($x_{Film}$/2 = $\sim$0.04). The maximum $T_c^{onset}$ position for the (Ba$_{1-x}$La$_x$)Fe$_2$As$_2$ films was $x_{Film}$/2 = $\sim$0.07, which is very close to that reported for the direct-doped Ba(Fe$_{1-x}$Co$_x$)$_2$As$_2$ single crystals (optimal doping concentration $x$ = $\sim$0.07). It is noteworthy that the dome widths for the indirect-doped (Ba$_{1-x}$La$_x$)Fe$_2$As$_2$ and the direct-doped Ba(Fe$_{1-x}$Co$_x$)$_2$As$_2$ are almost the same. In addition, the suppression rate of $T_{anom}$ (i.e., |d$T_{anom}$ / d$x_{Film}$|) shows close agreement with that of Ba(Fe$_{1-x}$Co$_x$)$_2$As$_2$. It is noted that the variations of $T_{anom}$ and $T_c^{onset}$ on the doping concentration closely overlap between the indirect-doped (Ba$_{1-x}$La$_x$)Fe$_2$As$_2$ and the direct-doped Ba(Fe$_{1-x}$Co$_x$)$_2$As$_2$. These results show a sharp contrast with those of 1111-type iron pnictides, in which the $T_c$ of indirect-doped SmFeAs(O$_{1-x}$F$_x$) (55 K),[2] is far higher than that of direct-doped Sm(Fe$_{1-x}$Co$_x$)AsO (17 K)[5] and the superconductivity dome width of $T_c$ for direct doping is much narrower than that of indirect doping[4,5].

We consider that there are two plausible reasons for the observed prominent differences between the 1111-type compounds and the 122-type compounds. One is the different dimensionality of their electronic structures, i.e., the Fermi surfaces of the 1111-type compounds are highly two dimensional due to the thicker blocking layers formed by the three atomic *RE*O layers, while those of the 122-type compounds are more three dimensional[24,25]. Therefore, the wave functions are more extended to the *AE* layers and hence, upon substitution of the *AE* site, the superconducting properties are more affected in the 122-type compounds. Another possibility is the different pairing symmetries of the superconducting states, i.e., it is considered that the 1111-type compounds are attributed to the sign-reversal $s_{+-}$ wave state that is sensitive to the doping sites, while the above observations in $BaFe_2As_2$ are more consistent with the non-sign-reversal $s_{++}$ wave state[13]. However, further effort is needed to clarify which idea is the controlling factor.

It would also be important to compare the phase diagram of $(Ba_{1-x}La_x)Fe_2As_2$ with those of indirect electron-doped $(Ca_{1-x}RE_x)Fe_2As_2$ single crystals[17], because the latter's maximum $T_c^{onset}$ (~47 K) is much higher than that of direct Co-doped $[Ca(Fe_{1-x}Co_x)_2As_2]$ (~20 K)[26], which is totally different from the present $BaFe_2As_2$ case, in spite of similar material systems and electronic structures. However, we should notice that all of the *RE*-doped $CaFe_2As_2$ single crystals reported to date showed very small shielding volume fractions[17,18]. In addition, $(Ca_{1-x}Pr_x)Fe_2As_2$ single crystals have two superconducting phases with $T_c$ of 21 and 49 K, where the former would be the bulk $T_c$ of $(Ca_{1-x}Pr_x)Fe_2As_2$ because of a large diamagnetism while the latter with a small shielding volume fraction may be attributed to a trace secondary phase or a strained surface phase[18]. If one takes the bulk $T_c$ of 21 K for $(Ca_{1−x}RE_x)Fe_2As_2$, it is consistent with the present results.

In conclusion, indirect electron doping into $BaFe_2As_2$ was attained by using epitaxial films of La-doped $BaFe_2As_2$ on MgO (001) substrates. The nonequilibrium film-growth process is an effective way to stabilize the metastable doping of La at Ba sites. The La substitution rapidly suppressed the AFM transition and induced superconductivity at a maximum $T_c^{onset}$ of 22.4 K for $(Ba_{1−x}La_x)Fe_2As_2$. We obtained clear superconducting domes closely overlapping with those of direct electron-doped $Ba(Fe_{1−x}Co_x)_2As_2$. The identical effect of the indirect and the direct electron-doping clarified that $T_c$ does not depend on whether or not impurity dopants exist in the active pairing layers, but depends solely on doped carrier concentrations. This result is consistent with the stronger three dimensionality of the Fermi-surface electronic structure and the $s_{++}$ wave state in superconductivity of electron-doped $BaFe_2As_2$.

## Acknowledgment

This work was supported by the Japan Society for the Promotion of Science (JSPS), Japan, through the "Funding Program for World-Leading Innovative R&D on Science and Technology (FIRST Program)".

Footnote:

\*: Present address: Frontier Research Center, Tokyo Institute of Technology

†: Corresponding author: hosono@msl.titech.ac.jp

**Figure captions**

FIG. 1. (Color online) Variation in the unit cell dimension in La-doped $BaFe_2As_2$ thin films and bulk polycrystals with doping concentration. Data were taken at room temperature for bulk polycrystals (open symbols) and thin films (solid symbols). The doping concentrations are taken as the nominal values for bulk polycrystals ($x_{Bulk}$) and those measured by EPMA for thin films ($x_{Film}$).

FIG. 2. (Color online) Variation of $\rho$–$T$ curves with doping concentration in $(Ba_{1-x}La_x)Fe_2As_2$ epitaxial films ($x_{Film}$ = 0–0.44). Doping concentrations $x_{Film}$ are indicated in the figures. The right-hand panel shows an enlarged view around $T_c$ ($T \leq 35$ K). The arrows in the left- and right-hand panels indicate the positions of temperatures of resistivity anomaly ($T_{anom}$) and onset $T_c$ ($T_c^{onset}$), respectively.

FIG. 3. (Color online) Electromagnetic properties of optimally doped $(Ba_{1-x}La_x)Fe_2As_2$ epitaxial films with $x_{Film}$ = 0.13. (a) Temperature dependence of the dc susceptibility under a magnetic field $\mu_0H$ = 5 G applied parallel to the *c*-axis. (b) Transverse resistivity $\rho_{xy}$ under magnetic fields up to 9 T at $T$ = 300–25 K. The inset shows the $T$ dependence of Hall coefficient $R_H$ estimated from the high $\mu_0H$ region ($\geq 3$ T).

FIG. 4. (Color online) Electronic phase diagram of indirect electron-doped $(Ba_{1-x}La_x)Fe_2As_2$. Open and solid symbols indicate the $T_{anom}$ and $T_c^{onset}$, respectively. Those reported for directly electron-doped $Ba(Fe_{1-x}Co_x)_2As_2$ single crystals (Ref. 7) are also shown for comparison. The inset figure illustrates the concept of indirect and direct dopings in the $BaFe_2As_2$ crystal.

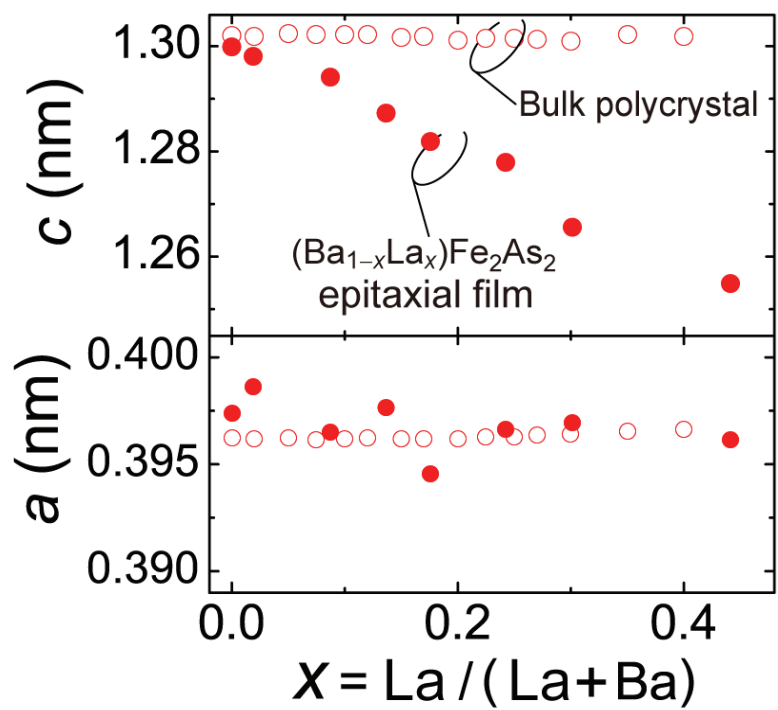

Figure 1

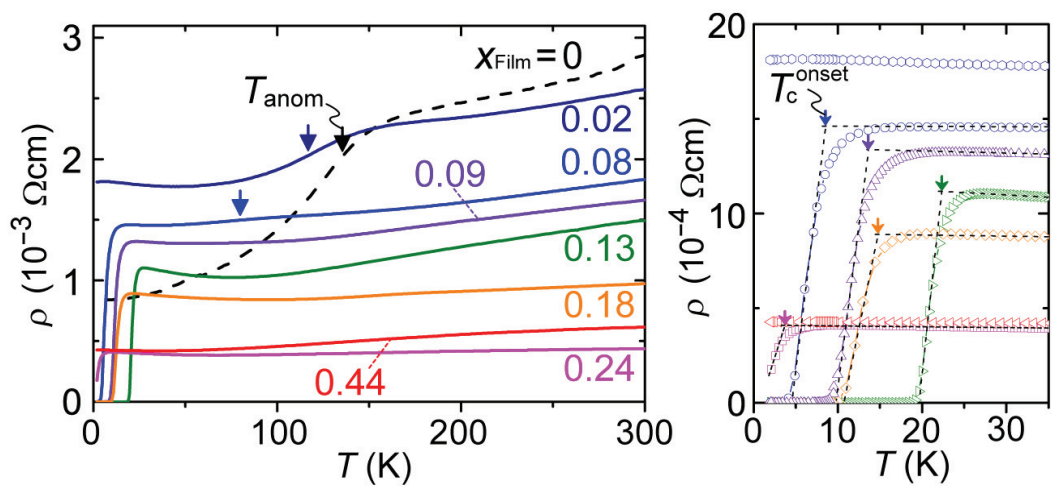

Figure 2

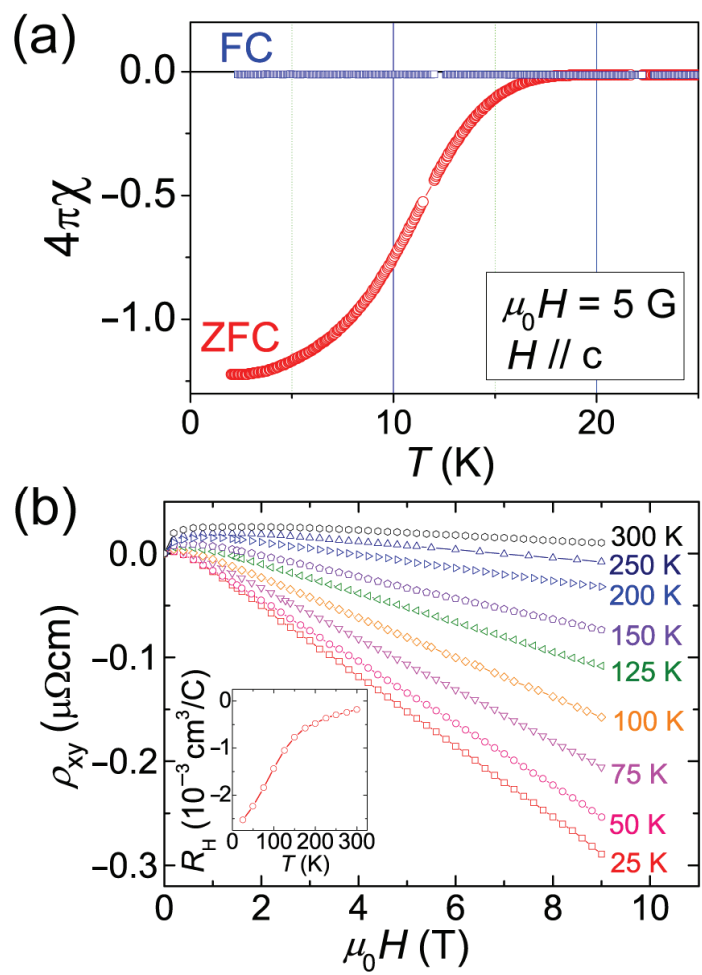

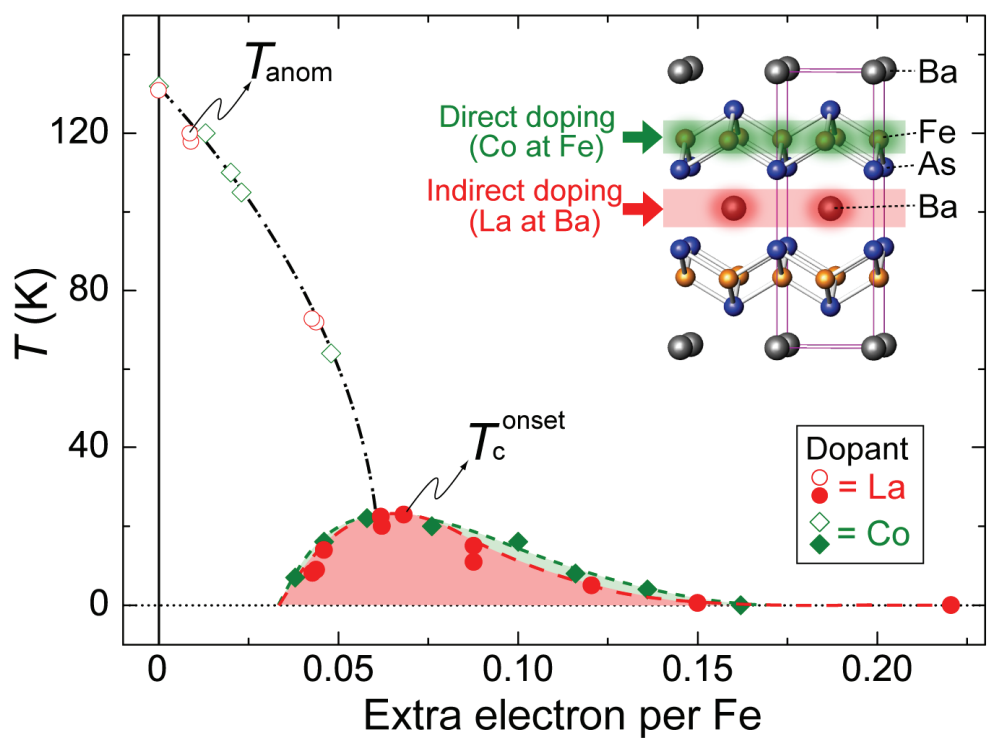

Figure 4

Supplementary Information for

**"Identical effects of indirect and direct electron doping of superconducting BaFe$_2$As$_2$ thin films"**

Takayoshi Katase[1], Soshi Iimura[1], Hidenori Hiramatsu[1], Toshio Kamiya[1], and Hideo Hosono[1,2,*]

[1] Materials and Structures Laboratory, Tokyo Institute of Technology, Mailbox R3-1, 4259 Nagatsuta-cho, Midori-ku, Yokohama 226-8503, Japan

[2] Frontier Research Center, Tokyo Institute of Technology, S2-6F East, Mailbox S2-13, 4259 Nagatsuta-cho, Midori-ku, Yokohama 226-8503, Japan

[(*)] E-mail: hosono@msl.titech.ac.jp

## I. Characterization of bulk polycrystals and epitaxial films of La-doped BaFe$_2$As$_2$

Figure S1 shows the powder x-ray diffraction (PXRD) patterns of the polycrystalline bulk samples of (Ba$_{1-x}$La$_x$)Fe$_2$As$_2$ ($x_{Bulk}$ = 0 − 0.4) synthesized by a solid-state reaction. These data indicate that the major phase is BaFe$_2$As$_2$ with a small amount of LaAs impurities and that the lattice parameters do not change by La adition.

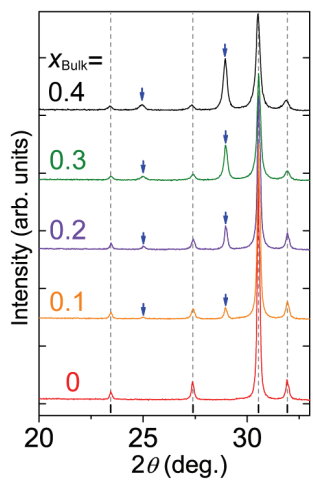

**Figure S1** PXRD patterns of (Ba$_{1-x}$La$_x$)Fe$_2$As$_2$ bulk polycrystals obtained by conventional solid-state reactions. Nominal compositions are $x_{Bulk}$ = 0 − 0.4. The arrows indicate LaAs impurity phases. The vertical dashed lines denote the simulated diffraction angles of the undoped BaFe$_2$As$_2$ phase.

Figure S2 shows the high-resolution XRD (HR-XRD) patterns of the (Ba$_{1-x}$La$_x$)Fe$_2$As$_2$ epitaxial films with $x_{Film}$ = 0 − 0.44 grown by pulsed laser deposition using the above (Ba$_{1-x}$La$_x$)Fe$_2$As$_2$ polycrystalline targets. The left panel shows the out-of-plane HR-XRD patterns. The 00*l* diffractions clearly shifted to higher angles from the diffraction peak of undoped BaFe$_2$As$_2$ films depending on the doping concentration $x_{Film}$. Epitaxial growth was confirmed by the in-plane $\phi$ scans of 200 diffractions shown in the right panels of Fig. S2. The $\phi$ scans exhibited a four-fold rotational symmetry originating from the tetragonal symmetry of the BaFe$_2$As$_2$ lattice.

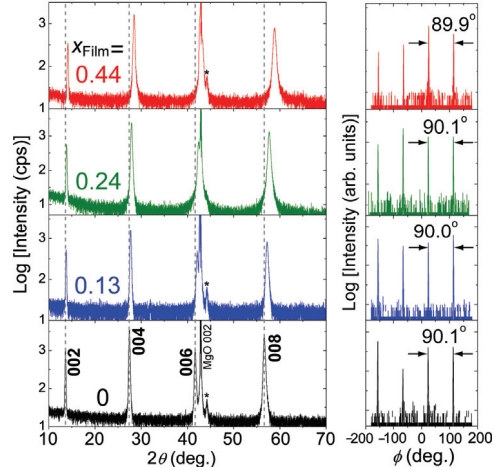

**Figure S2** HR-XRD patterns of (Ba$_{1-x}$La$_x$)Fe$_2$As$_2$ epitaxial films with $x_{Film}$ = 0 – 0.44 grown by PLD. (Left pannel) out-of-plane XRD patterns and (right pannel) the in-plane $\phi$ scans of 200 diffractions of each film. The asterisks indicate the difraction peaks of Fe impurity phase. The vertical dashed lines denote the 00*l* diffraction angles of the undoped BaFe$_2$As$_2$ phase.

Figure S3 shows the chemical composition mapping images of La for the (Ba$_{1-x}$La$_x$)Fe$_2$As$_2$ epitaxial films with $x_{Film}$ = 0.13 – 0.44, examined by electron-probe microanalysis (EPMA). The EPMA mapping images ensured the homogeneous distribution of La in the (Ba$_{1-x}$La$_x$)Fe$_2$As$_2$ epitaxial films.

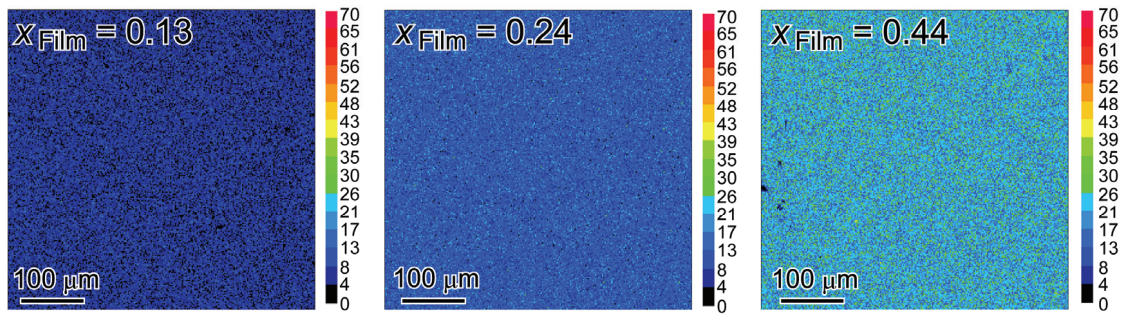

**Figure S3** EPMA mapping images of La concentration in (Ba$_{1-x}$La$_x$)Fe$_2$As$_2$ epitaxial films with $x_{Film}$ = 0.13, 0.24, and 0.44. The horizonal bars in the maps indicate the 100 μm scale, and the numbers beside the color scale bars indicate the measured signal intensity in count per pixel.